\documentclass[preprint,aps,prl,superscriptaddress,showpacs,citeautoscript]{revtex4-1}

\usepackage{graphicx}
\usepackage{dcolumn}
\usepackage{bm}
\usepackage{flafter}
\usepackage{lettrine}
\usepackage{amsmath}
\usepackage{amssymb}
\usepackage{color}
\usepackage{epsfig}

\usepackage{type1cm}
\usepackage{tabu}

\begin{document}

\title{Anharmonicity-induced isostructural phase transition of Zirconium under pressure. }

\author{Elissaios Stavrou}
\email{stavrou1@llnl.gov}
\affiliation {Lawrence Livermore National Laboratory, Physical and Life Sciences Directorate, P.O. Box 808 L-350, Livermore, California 94550, USA}
\author{Lin H. Yang}
\affiliation {Lawrence Livermore National Laboratory, Physical and Life Sciences Directorate, P.O. Box 808 L-350, Livermore, California 94550, USA}
\author{Per S\"{o}derlind}
\affiliation {Lawrence Livermore National Laboratory, Physical and Life Sciences Directorate, P.O. Box 808 L-350, Livermore, California 94550, USA}
\author{Daniel Aberg}
\affiliation {Lawrence Livermore National Laboratory, Physical and Life Sciences Directorate, P.O. Box 808 L-350, Livermore, California 94550, USA}
\author{Harry B. Radousky}
\affiliation {Lawrence Livermore National Laboratory, Physical and Life Sciences Directorate, P.O. Box 808 L-350, Livermore, California 94550, USA}
\author{Michael R. Armstrong}
\affiliation {Lawrence Livermore National Laboratory, Physical and Life Sciences Directorate, P.O. Box 808 L-350, Livermore, California 94550, USA}
\author{Jonathan L. Belof}
\affiliation {Lawrence Livermore National Laboratory, Physical and Life Sciences Directorate, P.O. Box 808 L-350, Livermore, California 94550, USA}
\author{Martin Kunz}
\affiliation{Advanced Light Source, Lawrence Berkeley Laboratory, Berkeley, California 94720, United States}
\author{Eran Greenberg}
\affiliation{Center for Advanced Radiation Sources, University of Chicago, Chicago, IL 60637, USA}
\author{Vitali B. Prakapenka}
\affiliation{Center for Advanced Radiation Sources, University of Chicago, Chicago, IL 60637, USA}
\author{David A. Young}
\email{young5@llnl.gov}
\affiliation {Lawrence Livermore National Laboratory, Physical and Life Sciences Directorate, P.O. Box 808 L-350, Livermore, California 94550, USA}

\begin{abstract}
We have performed a detailed x-ray diffraction structural study of Zr under pressure and unambiguously identify the existence of a first-order isostructural bcc-to-bcc phase transition near 58 GPa. First-principles quantum molecular dynamics lattice dynamics calculations support the existence of this phase transition, in excellent agreement with experimental results, triggered by anharmonic effects. Our results highlight the potential ubiquity of anharmonically driven isostructural transitions within the periodic table under pressure and calls for follow-up experimental and theoretical studies.

\end{abstract}
\maketitle

%\section{Introduction}
The interplay between pressure-induced structural phase transitions and electronic and lattice-dynamics properties under pressure is one of the most fundamental issues in condensed matter physics. In this context, pressure-induced first-order (\emph{i.e.} accompanied by a volume decrease) isostructural phase transitions of pure elements are extremely rare. Indeed, only one element, Ce, is known to have an isostructural fcc$\rightarrow$fcc phase transition under pressure \cite{Lawson1949} with a substantial volume decrease. Initial reports of a fcc$\rightarrow$fcc isostructural phase transition of Cs at 4.2 GPa \cite{Bridgman1927,Hall1964} are not supported by recent more detailed x-ray diffraction (XRD) studies \cite{Schwarz1998,McMahon2001}. Although the exact origin of the isostructural phase transition in Ce is still under debate \cite{Allen1982,Johansson1995,Krisch2011}, the general consensus is that this transition is triggered by a change in the degree of the localization and correlation of the one 4f electron of Ce \cite{Krisch2011}. This highlights the direct link between structural and electronic transition. Second-order isostructural phase transitions attributed to a change of compressibility and/or change of axial ratios have been reported for few elements under pressure, \emph{e.g.} in the case of Os \cite{Occeli2004,Dubrovinsky2015}; however, in that case no abrupt change of the cell volume was observed.

In 1991 an isostuctural bcc$\rightarrow$bcc phase transition in Zr was suggested by Akahama \emph{et al.} \cite{Akahama1991}, at $\sim$  57 GPa, using XRD. Akahama \emph{et al.} suggested that this transition is triggered by a s-d electronic transition, resembling the case of the, fcc$\rightarrow$fcc transition in Cs, according to the initial experimental report by Hall \emph{et al} \cite{Hall1964}. However, follow-up theoretical studies ruled out the possibility of such an electronic transition \cite{Jyoti1994}, didn't reproduce the experimental equation of state (EOS) above the claimed phase transition \cite{Ostanin1998}, and questioned even the existence of this isostructural transition \cite{Jyoti1994,Ostanin1998}, mainly due to the relatively low statistics of the experimental data points. Consequently, all recent relevant studies (\emph{e.g.} Ref. \cite{Lipp2017}) continue to list Ce as the only element with a first-order isostructural phase transition.

In this work we have performed a detailed experimental study of the structural evolution in Zr under pressure up to 210 GPa. Using state-of-the-art XRD under pressure that allowed us to record pressure dependent XRD patterns in intervals as low as 0.5 GPa, we report the existence a first-order isostuctural bcc$\rightarrow$bcc phase transition at $\sim$ 58 GPa realized by a major and abrupt decrease of the cell volume by 4\%. Moreover, we report a concomitant \emph{ab-initio} theoretical study aimed at explaining the experimentally observed phase transition. Since first-order isostructural transitions are very unusual in the elements, a theoretical explanation is needed. We have carried out a combined density-functional theory (DFT) and quantum molecular dynamics (QMD) calculation on bcc Zr in order to look for any anomaly that could explain the transition. We do not observe any subsequent structural phase transition up to the highest pressure of this study, \emph{i.e.} Zr remains in the bcc structure up to 210 GPa.

%\section{EXPERIMENTAL AND COMPUTATIONAL METHODS}
High-purity commercially available (Sigma-Aldrich $>$99.5\%) fine powder of Zr was loaded in a diamond anvil cell (DAC). Small quantities of ruby and gold powder were also loaded, for determination of pressure through ruby luminescence \cite{Syassen2008} and gold EOS, respectively. MAR-CCD detectors were used to collect pressure-dependent X-ray diffraction data at the undulator XRD beamline at GeoSoilEnviroCARS (sector13), APS, Chicago, and at the Advanced Light Source, Lawrence Berkeley National Laboratory Beamline 12.2.2. The X-ray probing beam spot size was focused to about 2-4$\mu$m at GeoSoilEnviroCARS and to 10 x 10$\mu$m at beamline 12.2.2 using Kirkpatrick-Baez mirrors. More details on the XRD experimental setups are given in Prakapenka et al. \cite{Prakapenka2008} and Kunz et al. \cite{Kunz2005}. Integration of powder diffraction patterns to yield scattering intensity versus 2$\theta$ diagrams and initial analysis were performed using the DIOPTAS \cite{Prescher2015} program. Le Bail refinements were performed using the GSAS \cite{Larson2000} software.

Our DFT calculations use the generalized gradient approximation (GGA) for the exchange-correlation functional \cite{Perdew1992}.  Our implementation is based on the full-potential linear muffin-tin orbitals (FPLMTO) method that has recently been described in detail \cite{Wills2000}.   In addition to the choice of GGA, we have found that no geometrical approximations (full potential), full relativity including spin-orbit coupling, and a well-converged basis set are often needed for good accuracy. Specifically for Zr, we associate a set of semi-core states 4\emph{s} and 4\emph{p} and valence states 5\emph{s}, 5\emph{p}, 4\emph{d}, and 4\emph{f}. Phonon dispersion and density-of-states were calculated for a set of volumes using the finite-displacement method implemented in the Phonopy code \cite{Togo2015}.  For this purpose, total energy calculations were performed within DFT.

In quantum molecular dynamic (QMD) simulations the ions move according to Newton's classical equations of motion in which the forces acting on the ions are computed \textquotedblleft on the fly\textquotedblright by solving the density-functional theory quantum-mechanical equations for the electrons at each discrete time step. Newton's time-dependent equation is discretized using a Verlet leap-frog algorithm. We use Born-Oppenheimer MD where the low-lying single-particle electronic eigenstates are computed by solving the self-consistent DFT Kohn-Sham \cite{Hohenberg1964,Kohn1965} equations within the framework of Mermin's finite temperature DFT \cite{Mermin1965}.

For the isothermal QMD simulations,  we have used a plane-wave pseudopotential method \cite{Hohenberg1964,Kohn1965,Yang2007} as implemented with optimized norm-conserving Vanderbilt (ONCV) pseudopotentials of Hamann \cite{Hamann2013,Hamann2017}.  A dual-projector ONCV scalar-relativistic pseudopotential for Zr was constructed using Perdew-Burke-Ernzerhof (PBE) gradient density functional \cite{Perdew1996} to treat 12 valence (4s$^2$ 4p$^6$ 4d$^2$ 5s$^2$ ) electrons. The electronic eigenstates are thermally occupied according to the Fermi-Dirac distribution function at a temperature T electron equivalent to the ion temperature. The use of a pseudopotential approximation and a plane-wave basis allows us to accurately calculate the forces acting on the ions. We perform all of our QMD simulations with a time step of 1.2 fs in a NVT ensemble with a constant number of particles (125 atoms) in which the volume is held constant within a fixed-shape simulation cell and the temperature is controlled and kept constant.  To test the convergence of our electronic eigenstates with this time step, we perform a single constant-energy, NVE-ensemble QMD simulation for 1.2 ps. The calculated energy fluctuations are less than 1 mRy per atom.

%\section{Results and discussion}
Fig. S1 of the Supplemental Material \cite{supp} shows integrated diffraction patterns of Zr at selected pressures above 40 GPa. The evolution of the XRD patterns doesn't show any discontinuities, appearance or disappearance of Bragg peaks, up to 210 GPa, thus revealing the stability of the bcc structure up to the highest pressure of this study. To determine the structural parameters, the diffraction patterns were analyzed by performing Le Bail refinements. We have in this way obtained the lattice parameters and the volume per atom and the results are shown in Figure 1. Structural evolution and EOS of Zr at lower pressures are in agreement with previous studies \emph{i.e.} we observe both the hcp$\rightarrow$$\omega$ and $\omega$$\rightarrow$bcc phase transitions at critical pressures that are in agreement with previous studies \cite{Jamieson1963,Xia1990,Akahama1991}. Our pressure-volume data are in agreement with those of Xia \emph{et al.} but we observe a lower volume at a given pressure in comparison to Akahama \emph{et al.}, especially at higher pressures. We believe that this can be attributed to the \textquotedblleft sandwich\textquotedblright-like Au-Zr-Au sample loading in Akahama \emph{et al.}. At pressures above 35 GPa the 111 Bragg peak of fcc Au coincides with the 110 Bragg peak of bcc Zr  (see Fig.1 in Ref.\cite{Akahama1991}), making pressure and EOS determination problematic.

\begin{figure}[ht]
\centering
\includegraphics[width=\linewidth]{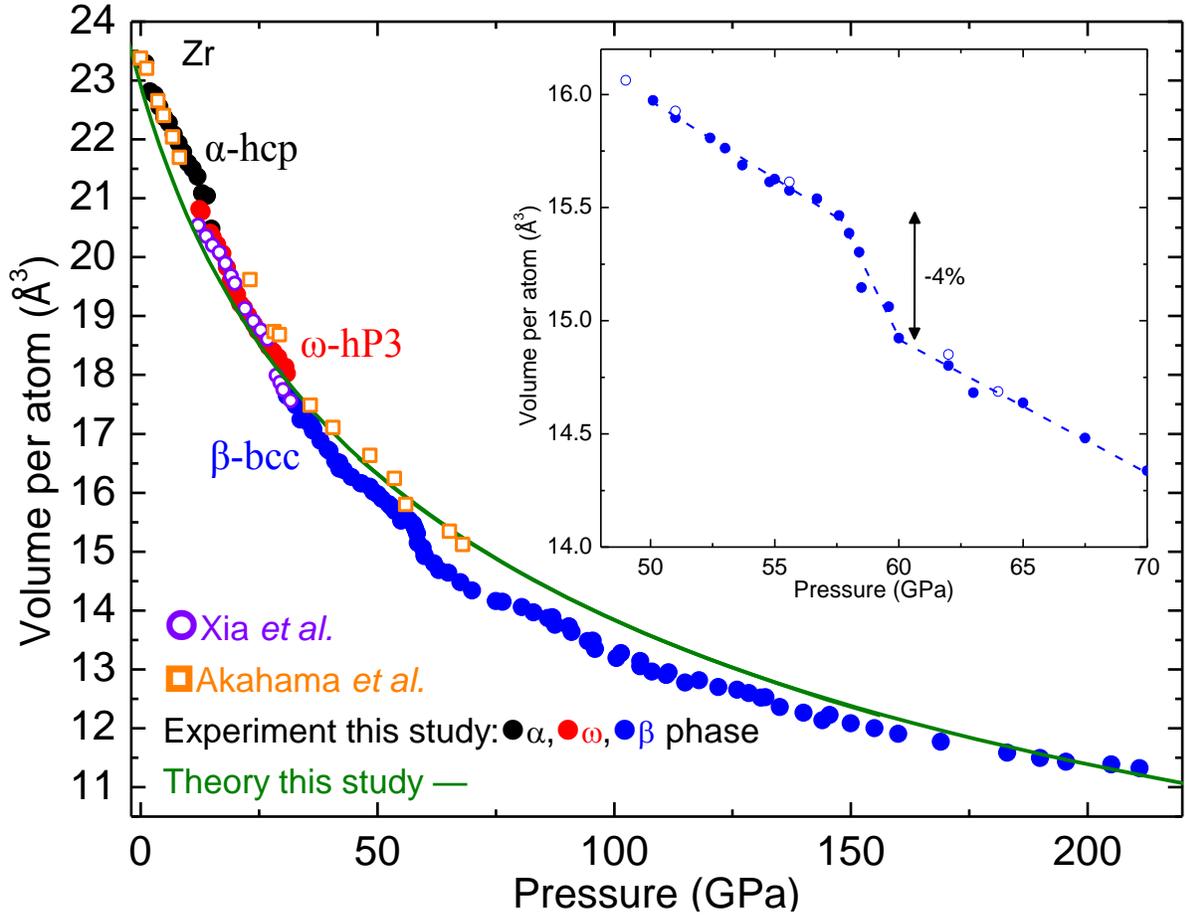}
\caption{ a) Volume-pressure data for the $\alpha$, $\omega$ and $\beta$ phases of Zr of this study together with the results from Refs \cite{Xia1990,Akahama1991}. The inset shows the volume-pressure data for the  $\beta$ phases of Zr in the range of the isostructural phase transition above 40 GPa on upstroke (closed circles)  and for the downstroke (open circles). The dashed lines are guides for the eye. }
\end{figure}

A major volume discontinuity is clearly observed in the 50-70 GPa pressure range, see inset of Fig. 1.   The volume per atom of the bcc structure decreases abruptly by -4\% between 57.5 and 60 GPa, a volume change that supports a pressure-induced first-order isostructural $\beta$$\rightarrow$$\beta'$ phase transition. Aiming to exclude the possibility of a subtle structural transition, \emph{e.g.} as in the case of the bcc to rhombohedral (bcc distortion) transition of vanadium at similar pressure \cite{Ding2007,Lee2007}, in Fig. 2 we plot the d-spacings and the widths at half maximum of the first (in increasing 2$\theta$) five observed bcc Bragg peaks as a function of pressure. As can be clearly seen, the d-spacings of all observed Bragg peaks follow the same trend upon pressure increase. Moreover, the width of all observed bcc Bragg peaks remains practically constant throughout the phase transition thus ruling out the possibility of a distortion of the bcc structure \cite{Jenei2011}.

%The absence of a  pressure induced distortion of the BCC structure is further justified by  the calculated enthalpies as a function of the rhombohedral distortion $\delta$ for various pressures, see Ref. \cite{Lee2007} for details. As it can be clearly  seen in figure 4(a), the rhombohedral angle value with the minimum enthalpy remains constant at $\alpha$=109.47$^o$ \emph{i.e.} the rhombohedral angle for the ideal BCC phase.

\begin{figure}[ht]
\centering
\includegraphics[width=\linewidth]{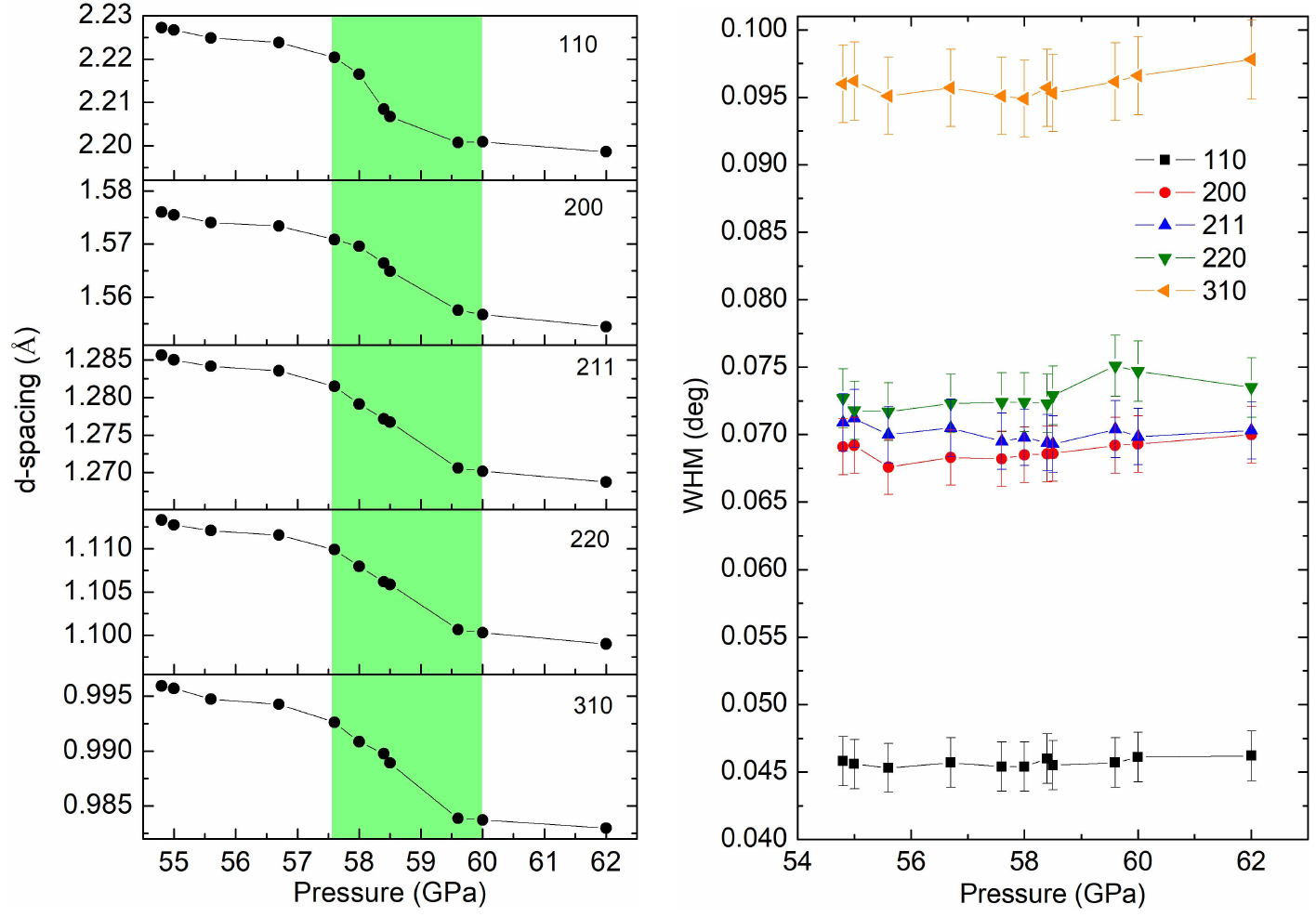}
\caption{Bragg peaks d-spacings (left) and widths at half maximum (right) \emph{vs} pressure of the $\beta$ phase of Zr in the pressure range of the isostructural phase transition. Corresponding Miller  indices are also noted.}
\end{figure}

Experimental limitations precluded a detailed determination of the EOS upon pressure release starting from pressures well above (65 GPa) the isostructural phase transition. Nevertheless, from the inset of Fig. 1 it is clear that the isostructural transition is fully reversible with no observable hysteresis to within 2 GPa. Moreover, in order to rule out any possibility of $\beta'$ bcc being a metastable phase with the parallel presence of a stable superstructure \cite{Hanfland2000} we have performed a prolonged (+40h) annealing of Zr at + 200 $^o$C at pressures just above the phase transition. Only an apparent decrease of the widths of the Bragg peaks was observed, without any sign of additional Bragg peaks that would indicate a superstructure \cite{Loa2016}.

We have fit two independent third-order Birch-Murnaghan EOSs \cite{Birch1978} to the experimental pressure-volume data below 57 GPa (low-pressure $\beta$ phase) and above 62 GPa (high-pressure $\beta'$ phase), respectively. The determined bulk modulus $B$ and its first derivative $B'$ at the experimental onset pressure for the two bcc phases are given in Table I. The results reveal a substantial decrease of the compressibility after the isostructural phase transition, $B$=142 GPa \emph{vs} $B$=255 GPa GPa for $\beta$-bcc and $\beta'$-bcc, respectively. On the other hand, first derivatives of bulk modulus with pressure, $B'$, remain practically the same and thus, highlighting the isostructural nature of the phase transition \cite{Stavrou2015}.

\begin{table}[tb]\centering
\footnotesize
\caption{Experimentally determined bulk modulus $B$, pressure derivative $B'$ and atomic volume at the experimental onset pressure for the $\beta$ and $\beta'$ bcc phases of Zr.}
\medskip
\setlength{\extrarowheight}{1pt}
\begin{ruledtabular}
\begin{tabular}{cccc}

Phase& $B$(GPa)&  $B'$ & V (\AA$^3$/atom) \\\hline
$\beta$& 142(8)&5.4(10)&17.64\\
$\beta'$&255(10)&5.3(5)&14.8\\

\end{tabular}
\end{ruledtabular}
\end{table}

In searching for an explanation of the isostructural transition, we have made a brief theoretical study of high-pressure zirconium using density functional theory (DFT). The static pressure-volume bcc isotherm has been computed and compared with experiment (Fig. 1) and the agreement is good, showing that DFT is accurate for Zr. The DFT isotherm shows no anomaly due to band crossing in agreement with previous theoretical studies \cite{Jyoti1994,Ostanin1998}. DFT phonons have been calculated in the transition region. Phonon dispersion curves for nine volumes around 60 GPa were generated, and these were used to compute Gr\"{u}neisen $\gamma$ parameters for six selected phonons. The $\gamma$ values were found to be positive and smoothly varying with pressure at all volumes, which excludes a phase transition due to negative values. Thus, a phase transition driven by harmonic lattice dynamics is unlikely. Another approach to lattice stability is to compute the elastic constants C$_{44}$ and C'= (C$_{11}$-C$_{12}$)/2 by distorting the lattice and finding the distortion energy \cite{Soderlind1993}. These constants do not show indications of instability in the transition pressure region; see Fig. S2. We also find that bcc Zr does not support the formation of magnetic moments at high pressure.

Our next step was to perform a more comprehensive theoretical approach involving first-principles anharmonic lattice dynamics using QMD. Molecular dynamics calculations \cite{Zhang1995} and inelastic neutron scattering measurements \cite{dubos1998} of the high-temperature ambient pressure bcc phase of Zr revealed a strong anharmonic dynamical behavior not predicted by traditional phonon theory and accompanied by a strong asymmetric scattering cross section. Thus, anharmonicity might also have an effect on the high-pressure bcc phase even at room temperature.

The first set of simulations was performed at 300 K, and the isothermal EOS is shown is Fig. 3 (black circles). The EOS agrees well with experimental data below 55 GPa and shows no sign of phase transitions. The second set of simulation were performed at 1000 K for about 10 picoseconds ($\sim$50 vibrations) and then  cooled down to 300 K by a cooling rate of 100 K/ps. At 300 K, after the relaxation process, we continue to run the simulations for another 10 picoseconds to calculate the average pressure and temperature. This heating and cooling process aims to permute the anharmonic vibrational modes in the system so that the Zr system achieves equipartition.  We note that the observation of relatively long timescales to equilibration for anharmonic systems is well-known\cite{Parisi1997} (indeed, even goes back to some of the first computer simulations ever performed \cite{Fermi1955,Livi1985}) and the requirement for an accelerated equilibration process to reveal the phase transition within the 10 ps quantum dynamics here is commensurate with these considerations. As shown in Fig. 3 of the P-V plot (red circles), the crystal system undergoes a first order phase transition above 54 GPa, with a volume decrease in excellent agreement with experimental results and a negligible (3 GPa, see inset of Fig. 4)) critical pressure difference.

\begin{figure}[ht]
\centering
\includegraphics[width=\linewidth]{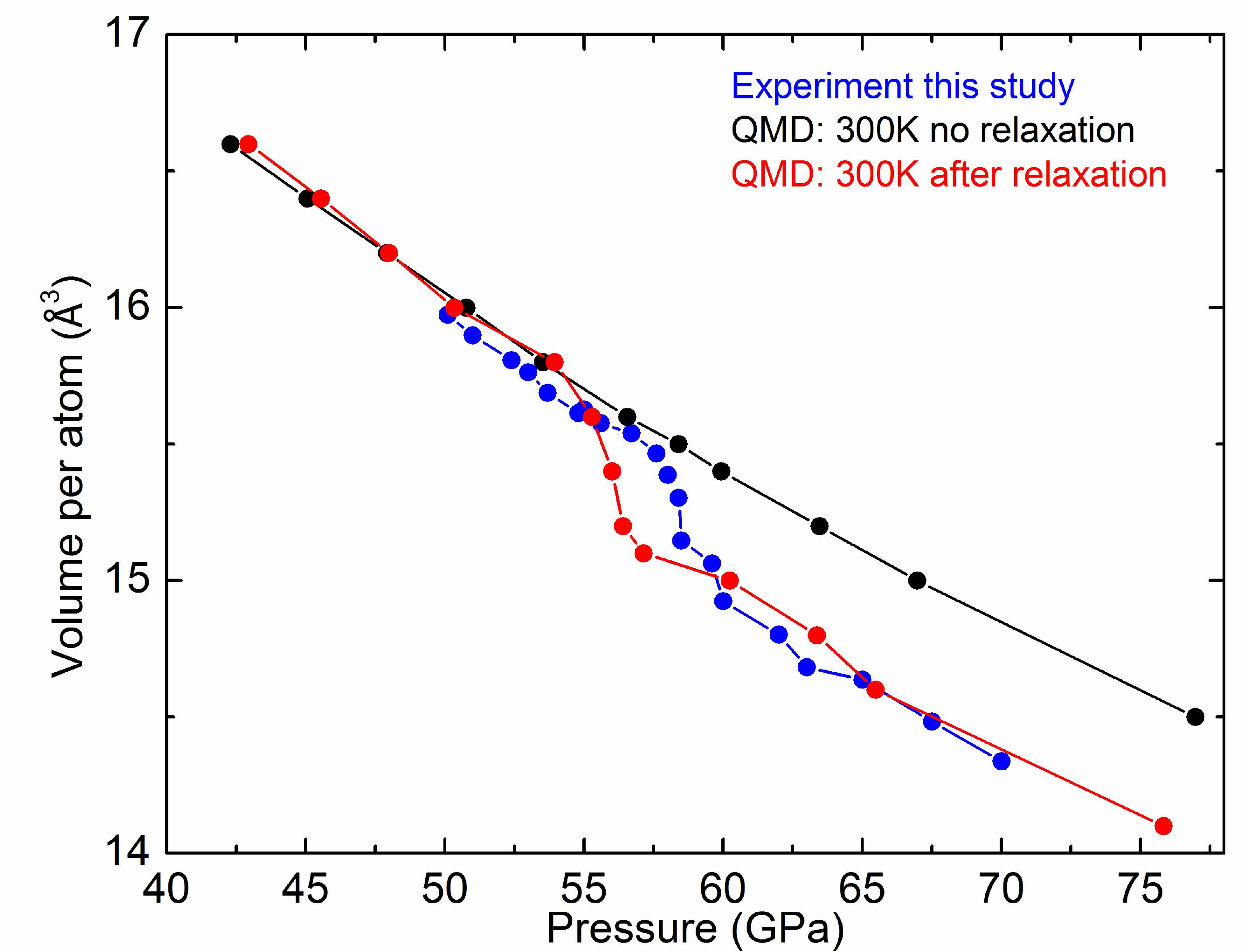}
\caption{Volume-pressure data for the  bcc phases of Zr in the range of the isostructural phase transition above 40 GPa. Experimental and QMD calculated values are shown with solid blue and black (300 K) and red (300 K after relaxation), respectively.}
\end{figure}

In order to determine the phase boundary between   $\beta$ and  $\beta'$, we have performed additional isochoric simulations for temperatures range between 100 K and 1000 K with a 100 K spacing.  Our results suggest that the $\beta$ and  $\beta'$ boundary line appears between 200-700 K with a slope of 166 K/GPa. The QMD EOS at 1000 K shows no sign of an isostructural phase transition and the  appearance of the phase transition ends at a temperature and pressure near 683 K and  51 GPa, respectively,  where the volume change goes to zero, see inset of Fig. 4.  This represents a critical point, instead of a triple point on the melting curve, resembling the case of cerium between the alpha and gamma phases.  Above the critical point, the two phases merge into a common bcc phase. At 100 and 200 K no transition was observed in QMD calculations, implying the absence of anharmonic effects at low temperatures.  In the inset of Fig.4 we provide a tentative P-T phase diagram of Zr according to the QMD results of this study. From the phase diagram is clear that the  Hugoniot is not  crossing the $\beta$-$\beta'$ phase boundary and this explains why previous gas-gun experiments didn't report the presence of this phase transition \cite{Greef2005}. However,  it is plausible to assume that ramp experiment  along the principle isentrope would probably be able to probe it.

This transition is attributed to anharmonic effects as can be seen in the calculated fractional thermal displacement function shown in Fig. 4. The thermal displacements $<R>$ as a function of pressure were calculated by averaging over 10 ps of QMD simulations for each isochore (pressure). The thermal displacement of each atom at each time step was calculated by taking the displacement of each atom from its ideal bcc atomic positions at T=0. The average of this quantity over the 10 ps QMD duration is plotted in Fig. 4 as a fraction of the nearest-neighbor interatomic distance for the bcc lattice.  The increased value of the function for the high-pressure phase indicates strong anharmonicity.

\begin{figure}[ht]
\centering
\includegraphics[width=\linewidth]{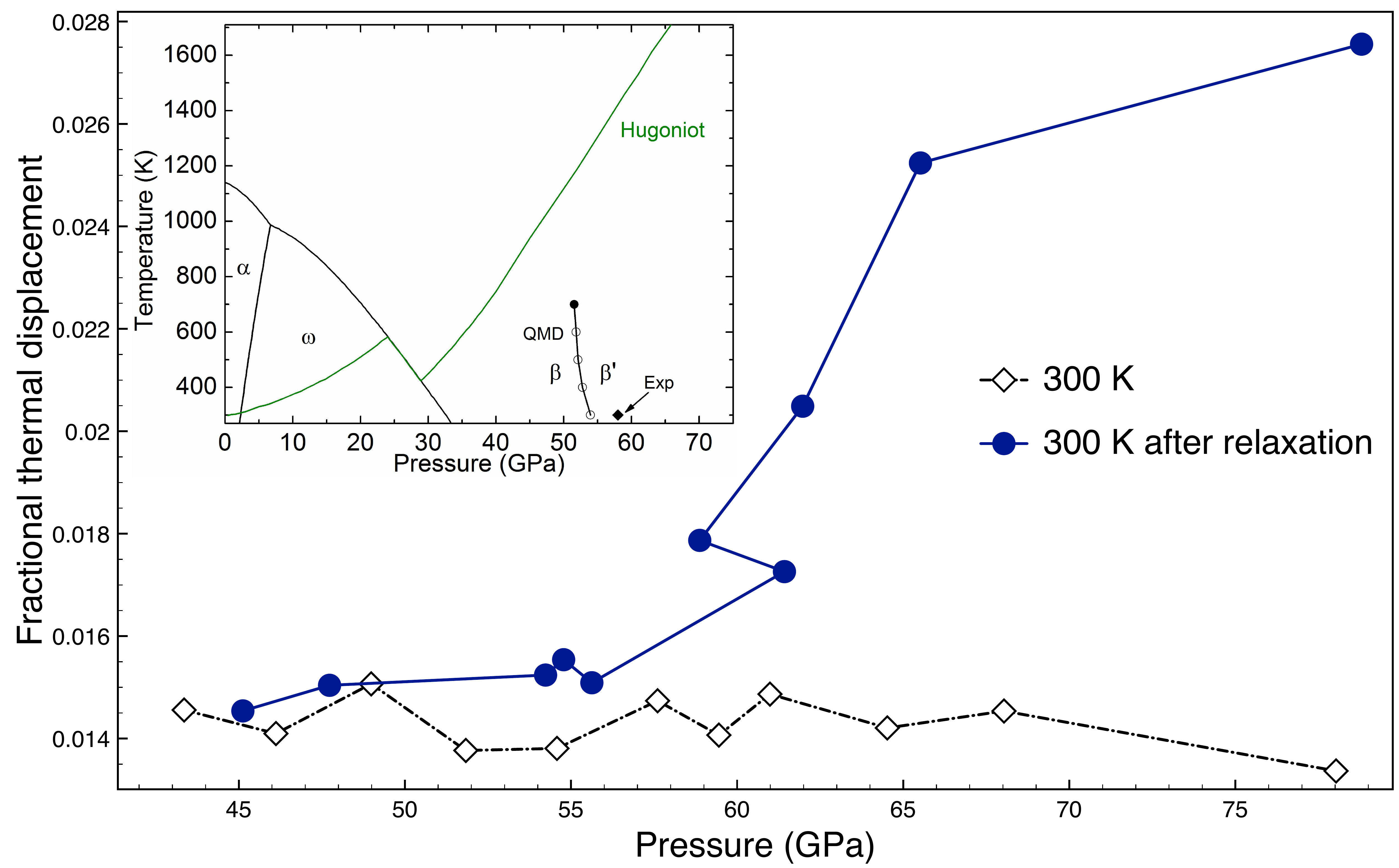}
\caption{ Calculated factional thermal displacements as a function of pressure for the relaxed and unrelaxed QMD simulations at 300K. The inset shows an updated Pressure-Temperature phase diagram of Zr. Shock Hugoniot (green line) and $\alpha$, $\omega$ and $\beta$ phase boundaries as reported  by Greeff \cite{Greef2005}. The phase boundary between $\beta$ and $\beta'$ phases as determined by QMD calculations is shown with open symbols. The solid circle at 700 K denotes the critical point. The experimentally determined critical pressure at RT is shown with a solid rhombus. }
\end{figure}

%\section{Summary}
In summary, we have performed a detailed XRD structural study of Zr under pressure and unambiguously identify the existence of a first order isostructural bcc to bcc ($\beta$$\rightarrow$$\beta'$) phase transition. Conventional T= 0 K DFT calculations could not provide an explanation for the origin of this phase transition based on either electronic or harmonic lattice dynamics properties. On the other hand, first-principles finite temperature QMD simulations unambiguously support the idea of a first-order pressure-induced isostructural phase transition triggered by anharmonic motion. Our study opens a new chapter in the structural behavior of elements under pressure by adding Zr, together with Ce, to the extremely narrow list of elements with pressure-induced first-order isostructural phase transitions. However, in contrast to the case of Ce, neither an electronic transition nor harmonic lattice dynamics properties can explain this transition. The results of our QMD calculations highlight that additional mechanisms, such as anharmonic motion, can be involved in such transitions \cite{Trubitsin2008}. Our study calls for follow-up experimental and theoretical studies aiming to explore similar first-order phase transitions in other transition metal elements and to fully elucidate the mechanism of the observed phase transition. We argue that such transitions might be possible in other relevant elements; however, the low precision of the P-V measurements in the past may have precluded the observation of such first-order transitions.

\acknowledgments
This work was performed under the auspices of the U. S. Department of Energy by Lawrence Livermore National Security, LLC under Contract DE-AC52-07NA27344. We gratefully acknowledge the LLNL LDRD program for funding support of this project under 16-ERD-037. GSECARS is supported by the U.S. NSF (EAR-1128799) and DOE Geosciences (DE-FG02-94ER14466).  The ALS is supported by the Director, Office of Science, BES of DOE under Contract No. DE-AC02-05CH11231, DE-AC02-06CH11357. E.S. thanks K. Syassen for fruitful discussions and for a critical reading of the manuscript. E.S. was partially supported by  LDRD grant 18-LW-036.

The authors declare no competing financial interests.

\end{document}